# Fundamental solutions for isotropic size-dependent couple stress elasticity

Ali R. Hadjesfandiari, Gary F. Dargush

ah@buffalo.edu, gdargush@buffalo.edu

Department of Mechanical and Aerospace Engineering, University at Buffalo, State University of New York, Buffalo, NY 14260 USA

**ABSTRACT**

Fundamental solutions for two- and three-dimensional linear isotropic size-dependent couple stress elasticity are derived, based upon the decomposition of displacement fields into dilatational and solenoidal components. While several fundamental solutions have appeared previously in the literature, the present version is for the newly developed fully determinate couple stress theory. Within this theory, the couple stress tensor is skew-symmetrical and thus possesses vectorial character. The present derivation provides solutions for infinite domains of elastic materials under the influence of unit concentrated forces and couples. Unlike all previous work, unique solutions for displacements, rotations, force-stresses and couple-stresses are established, along with the corresponding force-tractions and couple-tractions. These fundamental solutions are central in analysis methods based on Green's functions for infinite domains and are required as kernels in the corresponding boundary integral formulations for size-dependent couple stress elastic materials.

## 1. Introduction

*1.1 Background*

It has long been suggested that the strength of materials has size-dependency in smaller scales. Consequently, a number of theories that include either second gradients of deformation or extra rotational degrees of freedom, called microrotation, have been proposed to capture this size-effect. These developments have impacts at the finest continuum scales, such as micro- and nano-scales, which in turn can affect modern micro- and nano-technology.

The inclusion of the above new measures of deformation requires the introduction of couple-stresses in these materials, alongside the traditional force-stresses. The possible existence of couple-stress in materials was originally postulated by Voigt [1], while Cosserat and Cosserat [2] were the first to develop a mathematical model to analyze materials with couple-stresses. The Cosserats developed a general theory by considering microrotation, independent of the classical macrorotation. At first, experts in continuum mechanics did not see any necessity for the extra artificial degrees of freedom and considered the macrorotation vector as the sole degrees of freedom describing rotation. Toupin [3], Mindlin and Tiersten [4], Koiter [5] and others, following this line, considered the gradient of the rotation vector, as the curvature tensor, the effect of the second gradient of deformation in couple stress elastic materials. However, there are some difficulties with these formulations. The more notable ones relate to the indeterminacy of the spherical part of the couple-stress tensor and the appearance of the body couple in constitutive relations for the force-stress tensor [4]. For isotropic elastic materials, there are two size dependent elastic constants required in this theory. However, the linear equilibrium equations in terms of displacements inexplicably involve only one of these constants. This inconsistent theory usually is called the indeterminate couple stress theory in the literature (Eringen [6]) and clearly cannot be used as a basis for reliable prediction of material behavior.

As a result of the inconsistencies noted above, a number of alternative theories have been developed. One branch revives the idea of microrotation and is called micropolar theories (e.g., Mindlin [7]; Eringen [6]; Nowacki [8]). However, these theories exhibit inconsistencies, because microrotation is not a true continuum mechanical concept. The effect of the discontinuous microstructure of matter cannot be represented mathematically by an artificial continuous microrotation. The other main branch, labeled second gradient theories, avoids the idea of microrotation by introducing gradients of strain or rotation (e.g., Mindlin and Eshel [9]; Lazar et al. [10]). Although these theories use true continuum representations of deformation, the resulting formulations are not consistent with correct boundary condition specifications and energy conjugacy requirements.

Recently, the present authors [11] have resolved all of the above difficulties and have developed the consistent couple stress theory for solids. It has been shown that a consistent size-dependent continuum mechanics should involve only true continuum kinematical quantities without recourse to any additional artificial degrees of freedom. By using the definition of admissible boundary conditions, the principle of virtual work and some kinematical considerations, we have shown that the couple-stress tensor has a vectorial character and that the body couple is not distinguishable from the body force. The work also demonstrates that the stresses are fully determinate and the measure of deformation corresponding to couple-stress is the skew-symmetrical mean curvature tensor. This development can be extended quite naturally into many branches of continuum mechanics, including, for example, elastoplasticity and piezoelectricity. However, the first step is the development of infinitesimal linear isotropic elasticity, which involves only a single size-dependent constant.

Since this theory is much more complicated than Cauchy elasticity, analytical solutions are rare and, consequently, numerical formulations are needed to solve more general size-dependent couple stress elastic boundary value problems. Interestingly, it seems the boundary element method is a suitable numerical tool to solve a wide range of couple stress elastic boundary value problems. However, this requires the free space Green's functions or fundamental solutions as the required kernels to transform the governing equations to a set of boundary integral equations. These fundamental solutions are the elastic solutions for an infinitely extended domain under the influence of unit concentrated forces and couples. The importance of these fundamental solutions is enhanced further, when we notice that the free space Green's functions play a direct role in the solution of many practical problems for infinite domains.

Mindlin and Tiersten [4] have given the necessary potential functions for obtaining the three-dimensional displacement fundamental solutions in the indeterminate isotropic elasticity. Chowdhury and Glockner [12] provide analogous functions by a matrix inversion technique for steady state vibration. Interestingly, we find that our linear equilibrium equations in terms of displacements in the determinate theory are identical to those of Mindlin and Tiersten [4]. This means the above displacement solutions are valid in our determinate theory. However, due to the indeterminacy of stresses, appearance of body couples independent of body force and existence of two size-dependent elastic constants, the corresponding stresses were not obtainable within previous couple stress theory.

In the present work, we obtain all two- and three-dimensional displacement and stress fundamental solutions for our consistent isotropic couple stress elasticity. Within this theory, everything is fully determinate and depends on only a single size-dependent material constant. For the three-dimensional case, we derive the displacement kernels directly by a decomposition method and then determine the corresponding stresses.

Two-dimensional stress fundamental solutions have been presented in Huilgol [13] only for concentrated force, based on the indeterminate couple-stress theory developed by Mindlin [14,15]. Reference [11] shows that some of these developments for plane problems remain useful in our determinate couple stress elasticity. However, we should remember in Mindlin's theory, there is an extra material constant, along with indeterminacy in the spherical part of the couple-stress tensor. Here we derive the complete two-dimensional fundamental solutions for point force and point couple, including displacements, force- and couple-stresses with a method similar to that used in the three-dimensional case.

Before continuing with the development for couple stress elasticity, we should mention the work by Hirashima and Tomisawa [16], Sandru [17] and Khan et al. [18] to develop fundamental solutions within the framework of micropolar elasticity. Although their results have some similarities to our fundamental solutions, we should emphasize once again that microrotations are not a fully consistent continuum mechanics concept. Consequently, these solutions are of limited practical utility.

*1.2 Basic Equations*

Let us assume the three-dimensional coordinate system $x_1 x_2 x_3$ as the reference frame with unit vectors $\mathbf{e}_1$, $\mathbf{e}_2$ and $\mathbf{e}_3$. Consider an arbitrary part of the material continuum occupying a volume $V$ enclosed by boundary surface $S$. In a continuum mechanical theory for size-dependent couple stress materials, the equations of equilibrium become

$$\sigma_{ji,j} + F_i = 0 \tag{1}$$

$$\mu_{ji,j} + \varepsilon_{ijk}\sigma_{jk} = 0 \tag{2}$$

where $\sigma_{ji}$ and $\mu_{ji}$ are force- and couple-stress tensors, and $F_i$ is the body force per unit volume of the body. Within this theory, the force-stress tensor is generally non-symmetric and can be decomposed as

$$\sigma_{ji} = \sigma_{(ji)} + \sigma_{[ji]} \tag{3}$$

where $\sigma_{(ji)}$ and $\sigma_{[ji]}$ are the symmetric and skew-symmetric parts, respectively. In [11], we have shown that in a continuum mechanical theory, the couple-stress tensor is skew-symmetrical. Thus,

$$\mu_{ji} = -\mu_{ij} \tag{4}$$

Therefore, the couple-stress vector $\mu_i$ dual to the tensor $\mu_{ij}$ can be defined by

$$\mu_i = \frac{1}{2}\varepsilon_{ijk}\mu_{kj} \tag{5}$$

where we also have

$$\varepsilon_{ijk}\mu_k = \mu_{ji} \tag{6}$$

Then, the angular equilibrium equation gives the skew-symmetric part of the force-stress tensor as

$$\sigma_{[ji]} = -\mu_{[i,j]} \tag{7}$$

We can consider the axial vector $s_i$ dual to $\sigma_{[ji]}$, where

$$s_i = \frac{1}{2}\varepsilon_{ijk}\sigma_{[j,k]} \tag{8}$$

which also satisfies

$$\varepsilon_{ijk}s_k = \sigma_{[ji]} \tag{9}$$

It is seen that by using Eqs. (7) and (8), we obtain

$$s_i = \frac{1}{2}\varepsilon_{ijk}\mu_{k,j} \tag{10}$$

which can be written in the vectorial form

$$\mathbf{s} = \frac{1}{2}\nabla\times\boldsymbol{\mu} \tag{11}$$

Interestingly, it is seen that

$$\nabla\bullet\mathbf{s} = 0 \tag{12}$$

The force-traction vector at a point on a surface with unit normal vector $n_i$ can be expressed as

$$t_i^{(n)} = \sigma_{ji}n_j \tag{13}$$

Similarly, the moment-traction vector can be written

$$m_i^{(n)} = \mu_{ji}n_j = \varepsilon_{ijk}n_j\mu_k \tag{14}$$

which can also be written in the vectorial form

$$\mathbf{m}^{(n)} = \mathbf{n}\times\boldsymbol{\mu} \tag{15}$$

Reference [11] shows that in couple stress materials, the body couple is not distinguishable from the body force. The body couple $C_i$ transforms into an equivalent body force $\frac{1}{2}\varepsilon_{ijk}C_{k,j}$ in the volume and a force-traction vector $\frac{1}{2}\varepsilon_{ijk}C_j n_k$ on the bounding surface. In vectorial form, this means

$$\mathbf{F} + \frac{1}{2}\nabla \times \mathbf{C} \to \mathbf{F} \quad \text{in} \quad V \tag{16}$$

and

$$\mathbf{t}^{(n)} + \frac{1}{2}\mathbf{C} \times \mathbf{n} \to \mathbf{t}^{(n)} \quad \text{on} \quad S \tag{17}$$

This is the first important result required for the development of the fully determinate couple stress theory.

In addition, there is a need to introduce the appropriate kinematical and constitutive relations. As an initial step, the displacement gradients are decomposed into symmetrical and skew-symmetrical components, such that

$$u_{i,j} = e_{ij} + \omega_{ij} \tag{18}$$

where

$$e_{ij} = u_{(i,j)} = \frac{1}{2}\left(u_{i,j} + u_{j,i}\right) \tag{19}$$

$$\omega_{ij} = u_{[i,j]} = \frac{1}{2}\left(u_{i,j} - u_{j,i}\right) \tag{20}$$

Since the rotation tensor $\omega_{ij}$ is skew-symmetrical, one can introduce a dual rotation vector, such that

$$\omega_i = \frac{1}{2}\varepsilon_{ijk}\omega_{kj} \tag{21}$$

In the usual infinitesimal Cauchy elasticity, only the symmetric strain tensor $e_{ij}$ contributes to the stored energy. However, in the size-dependent couple stress elastic theory, mean curvatures $\kappa_{ij}$ also play a role, where

$$\kappa_{ij} = \omega_{[i,j]} = \frac{1}{2}\left(\omega_{i,j} - \omega_{j,i}\right) \tag{22}$$

From Eq. (7), one can recognize that the mean curvature tensor is skew-symmetrical and thus can be rewritten in terms of a dual mean curvature vector [11], where

$$\kappa_i = \frac{1}{2}\varepsilon_{ijk}\kappa_{kj} \tag{23}$$

Reference [11] derives the general constitutive relations for an elastic material, when stored energy is expressed in terms of the symmetrical strain tensor $e_{ij}$ and the mean curvature vector $\kappa_i$. Interestingly, for linear isotropic elastic media, the following constitutive relations can be written for the force-stress and couple-stress, respectively

$$\sigma_{(ji)} = \lambda e_{kk}\delta_{ij} + 2\mu e_{ij} \tag{24}$$

$$\mu_i = -8\eta\kappa_i \tag{25}$$

and therefore

$$\sigma_{[ji]} = -\mu_{[i,j]} = 8\eta\kappa_{[j,i]} = 2\eta\varepsilon_{ijk}\nabla^2\omega_k \tag{26}$$

Here $\lambda$ and $\mu$ are the usual Lamé elastic moduli, while $\eta$ is the sole additional parameter that accounts for couple stress effects in an isotropic material. It is seen that these relations are similar to those in the indeterminate couple stress theory (Mindlin, Tiersten [4]; Koiter [5]), when the two size-dependent properties have the relation $\eta' = -\eta$. Thus, in [11], we have derived couple stress theory with only one single size-dependent constant in which all former troubles with indeterminacy disappear. There is no spherical indeterminacy and the second couple stress coefficient $\eta'$ depends on $\eta$, such that the couple-stress tensor becomes skew-symmetric. Interestingly, the ratio

$$\frac{\eta}{\mu} = l^2 \tag{27}$$

specifies a characteristic material length $l$, which is absent in Cauchy elasticity, but is fundamental to the small deformation size-dependent elasticity theory under consideration here. It should be also noticed that

$$\lambda = 2\mu\frac{\nu}{1-2\nu} \tag{28}$$

where $\nu$ represents the usual Poisson ratio. Therefore, Eq. (24) can be written as

$$\sigma_{(ji)} = 2\mu\left(\frac{\nu}{1-2\nu}e_{kk}\delta_{ij} + e_{ij}\right) \tag{29}$$

and the total stress tensor becomes

$$\sigma_{ji} = 2\mu\left(\frac{\nu}{1-2\nu}e_{kk}\delta_{ij} + e_{ij}\right) + 2\mu l^2 \varepsilon_{ijk}\nabla^2\omega_k \tag{30}$$

By using the relations in Eqs. (19) and (21), this tensor can be written in terms of displacements as

$$\sigma_{ji} = \mu\left(\frac{2\nu}{1-2\nu}u_{k,k}\delta_{ij} + u_{i,j} + u_{j,i}\right) - \eta\nabla^2\left(u_{i,j} - u_{j,i}\right) \tag{31}$$

After substituting Eq. (31) into Eq. (1), we can rewrite the governing differential equations in terms of the displacement and body force density fields as

$$\left(\lambda + \mu + \eta\nabla^2\right)u_{k,ki} + (\mu - \eta\nabla^2)\nabla^2 u_i + F_i = 0 \tag{32}$$

or in vectorial form as

$$\left(\lambda + \mu + \eta\nabla^2\right)\nabla(\nabla \bullet \mathbf{u}) + (\mu - \eta\nabla^2)\nabla^2\mathbf{u} + \mathbf{F} = 0 \tag{33}$$

Note that Mindlin and Tiersten [4] previously derived an identical form within the context of their indeterminate couple stress theory. However, recall that the stresses in the Mindlin-Tiersten formulation not only involve two parameters $\eta$ and $\eta'$, but also are indeterminate.

The general solution for the displacement in the indeterminate theory has been derived by Mindlin and Tiersten [4]. This can be written

$$\mathbf{u} = \mathbf{B} - l^2\nabla\nabla\bullet\mathbf{B} - \frac{1}{4(1-\nu)}\nabla\left[\mathbf{r}\bullet\left(1 - l^2\nabla^2\right)\mathbf{B} + B_0\right] \tag{34}$$

where the vector function $\mathbf{B}$ and scalar function $B_0$ satisfy the relations

$$\mu\left(1 - l^2\nabla^2\right)\nabla^2\mathbf{B} = -\mathbf{F} \tag{35}$$

$$\mu\nabla^2 B_0 = \mathbf{r}\bullet\mathbf{F} \tag{36}$$

Although, remarkably, these remain valid for the determinate couple stress theory, we instead use the direct decomposition method in the present work to derive the displacement fundamental solutions. Afterwards, we continue by developing the corresponding rotations, curvatures and fully-determinate force- and couple-stresses and tractions, which has previously not been possible. This, in turn, enables the formulation of new boundary integral representations and the development of boundary element methods to solve a broad range of couple stress elastostatic boundary value problems in both two- and three-dimensions.

## 2. Fundamental Solutions for Three-Dimensional Case

In this section, we derive the fundamental solutions for the three-dimensional case. As was mentioned, these are the elastic solutions of an infinitely extended domain under the influence of unit concentrated forces and couples.

*2.1 Concentrated Force*

Assume that in the infinitey extended material, there is a unit concentrated force at the origin in an arbitrary direction specified by the unit vector **a**. This concentrated force can be represented as a body force

$$\mathbf{F} = \delta^{(3)}(\mathbf{x})\mathbf{a} \tag{37}$$

where $\delta^{(3)}(\mathbf{x})$ is the Dirac delta function in three-dimensional space. It is known that

$$\delta^{(3)}(\mathbf{x}) = -\nabla^2\left(\frac{1}{4\pi r}\right) \tag{38}$$

By applying the vectorial relation

$$\nabla \times (\nabla \times \mathbf{G}) = \nabla(\nabla \bullet \mathbf{G}) - \nabla^2 \mathbf{G} \tag{39}$$

for the vector $-\dfrac{1}{4\pi r}\mathbf{a}$, we decompose the body force as

$$\mathbf{F} = -\nabla^2\left(\frac{1}{4\pi r}\right)\mathbf{a} = \nabla \times \left(\nabla \times \frac{\mathbf{a}}{4\pi r}\right) - \nabla\left(\nabla \bullet \frac{\mathbf{a}}{4\pi r}\right) \tag{40}$$

If we consider the decomposition of the resulting displacement $\mathbf{u}^F$ as

$$\mathbf{u}^F = \mathbf{u}^{(1)} + \mathbf{u}^{(2)} \tag{41}$$

where $\mathbf{u}^{(1)}$ and $\mathbf{u}^{(2)}$ are the dilatational and solenoidal part of displacement $\mathbf{u}^F$ satisfying

$$\nabla \times \mathbf{u}^{(1)} = 0 \tag{42}$$

$$\nabla \bullet \mathbf{u}^{(2)} = 0 \tag{43}$$

It is seen that

$$(\lambda + 2\mu)\nabla^2 \mathbf{u}^{(1)} = \nabla\left(\nabla \bullet \frac{\mathbf{a}}{4\pi r}\right) \tag{44}$$

$$\eta \nabla^2 \nabla^2 \mathbf{u}^{(2)} - \mu \nabla^2 \mathbf{u}^{(2)} = \nabla \times \left(\nabla \times \frac{\mathbf{a}}{4\pi r}\right) \tag{45}$$

If we introduce two vectors $\mathbf{A}^{(1)}$ and $\mathbf{A}^{(2)}$, such that

$$\mathbf{u}^{(1)} = \nabla\left(\nabla \bullet \mathbf{A}^{(1)}\right) \tag{46}$$

$$\mathbf{u}^{(2)} = \nabla \times \left(\nabla \times \mathbf{A}^{(2)}\right) = \nabla\left(\nabla \bullet \mathbf{A}^{(2)}\right) - \nabla^2 \mathbf{A}^{(2)} \tag{47}$$

It is seen that they satisfy the conditions in Eqs. (42) and (43), respectively, and therefore

$$(\lambda + 2\mu)\nabla^2 \mathbf{A}^{(1)} = \frac{\mathbf{a}}{4\pi r} \tag{48}$$

$$\eta \nabla^2 \nabla^2 \mathbf{A}^{(2)} - \mu \nabla^2 \mathbf{A}^{(2)} = \frac{\mathbf{a}}{4\pi r} \tag{49}$$

It is obvious that the solutions should be in the form

$$\mathbf{A}^{(1)} = \varphi \mathbf{a} \tag{50}$$

$$\mathbf{A}^{(2)} = \psi \mathbf{a} \tag{51}$$

where $\varphi$ and $\psi$ are scalar functions of $r$ having radial symmetry. Therefore,

$$\nabla^2 \varphi = \frac{1}{4\pi(\lambda + 2\mu)r} \tag{52}$$

$$\eta \nabla^2 \nabla^2 \psi - \mu \nabla^2 \psi = \frac{1}{4\pi r} \tag{53}$$

where $\nabla^2$ is the laplacian operator in three dimensions. Because of radial symmetry, it reduces to

$$\nabla^2 \rightarrow \frac{d^2}{dr^2} + \frac{2}{r}\frac{d}{dr} = \frac{1}{r^2}\frac{d}{d}\left(r^2\frac{d}{dr}\right) \qquad (54)$$

The regular solutions to the Eqs. (52) and (53) are

$$\varphi = \frac{1}{8\pi(\lambda + 2\mu)} r \qquad (55)$$

$$\psi = \frac{\eta}{4\pi\mu^2}\frac{e^{-r/l} - 1}{r} - \frac{1}{8\pi\mu}r \qquad (56)$$

Therefore, we have

$$\mathbf{A}^{(1)} = \frac{1}{8\pi(\lambda + 2\mu)} r\mathbf{a} \qquad (57)$$

$$\mathbf{A}^{(2)} = -\frac{l^2}{4\pi\mu}\frac{1 - e^{-r/l}}{r}\mathbf{a} - \frac{1}{8\pi\mu} r\mathbf{a} \qquad (58)$$

It is seen that from Eq. (46)

$$u_i^{(1)} = \frac{1}{8\pi(\lambda + 2\mu)}\frac{1}{r}\left(\delta_{iq} - \frac{x_i x_q}{r^2}\right) a_q \qquad (59)$$

By using the relation in Eq. (28), we have

$$\lambda + 2\mu = 2\mu\frac{1-\nu}{1-2\nu} \qquad (60)$$

and Eq. (59) can be written as

$$u_i^{(1)} = \frac{1-2\nu}{16\pi\mu(1-\nu)}\frac{1}{r}\left(\delta_{iq} - \frac{x_i x_q}{r^2}\right) a_q \qquad (61)$$

It is also seen that from Eq. (47)

$$\mathbf{u}^{(2)} = \nabla(\nabla \bullet \mathbf{A}^{(2)}) + \frac{1}{4\pi\mu}\frac{1 - e^{-r/l}}{r}\mathbf{a} \qquad (62)$$

Therefore

$$u_i^{(2)} = \frac{1}{4\pi\mu} \frac{l^2}{r^3} \left[ \left\{ \left(3 + 3\frac{r}{l} + \frac{r^2}{l^2}\right) e^{-r/l} - 3 \right\} \frac{x_i x_q}{r^2} + \left\{ 1 - \left(1 + \frac{r}{l} + \frac{r^2}{l^2}\right) e^{-r/l} \right\} \delta_{iq} \right] a_q$$
$$+ \frac{1}{8\pi\mu} \frac{1}{r} \left( \frac{x_i x_q}{r^2} + \delta_{iq} \right) a_q \tag{63}$$

and for the total displacement, we have

$$u_i^F = u_i^{(1)} + u_i^{(2)} = \frac{1}{16\pi\mu(1-\nu)} \frac{1}{r} \left[ (3-4\nu)\delta_{iq} + \frac{x_i x_q}{r^2} \right] a_q$$
$$+ \frac{1}{4\pi\mu} \frac{l^2}{r^3} \left[ \left\{ \left(3 + 3\frac{r}{l} + \frac{r^2}{l^2}\right) e^{-r/l} - 3 \right\} \frac{x_i x_q}{r^2} + \left\{ 1 - \left(1 + \frac{r}{l} + \frac{r^2}{l^2}\right) e^{-r/l} \right\} \delta_{iq} \right] a_q \tag{64}$$

It should be noticed that the first part of this relation is the displacement from Cauchy elasticity. However, in our determinate theory, we can derive determinate stresses. These force- and couple-stresses are developed in the following.

As a first step in that direction, we find that the gradient of displacement from Eq. (64) is

$$u_{i,j}^F = \frac{1}{16\pi\mu(1-\nu)} \frac{1}{r^2} \left[ -(3-4\nu)\frac{x_j}{r}\delta_{iq} + \frac{\delta_{ij} x_q + \delta_{jq} x_i}{r} - \frac{3x_i x_j x_q}{r^3} \right] a_q$$
$$+ \frac{1}{4\pi\mu} \frac{l^2}{r^4} \left[ \begin{array}{l} \left\{ 15 - \left(15 + 15\frac{r}{l} + 6\frac{r^2}{l^2} + \frac{r^3}{l^3}\right) e^{-r/l} \right\} \frac{x_i x_j x_q}{r^3} \\ + \left\{ \left(3 + 3\frac{r}{l} + \frac{r^2}{l^2}\right) e^{-r/l} - 3 \right\} \frac{\delta_{ij} x_q + \delta_{jq} x_i + \delta_{iq} x_j}{r} \end{array} \right] a_q \tag{65}$$
$$+ \frac{1}{4\pi\mu} \frac{1}{r^2} \left(1 + \frac{r}{l}\right) e^{-r/l} \frac{x_j}{r} \delta_{iq} a_q$$

Therefore, the strain tensor becomes

$$e_{ij}^{F} = u_{(i,j)}^{F}$$

$$= \frac{1}{16\pi\mu(1-\nu)}\frac{1}{r^{2}}\left[-(3-4\nu)\frac{x_{j}\delta_{iq}+x_{i}\delta_{jq}}{2r}+\frac{2\delta_{ij}x_{q}+\delta_{jq}x_{i}+x_{j}\delta_{iq}}{2r}-\frac{3x_{i}x_{j}x_{q}}{r^{3}}\right]a_{q}$$

$$+\frac{1}{4\pi\mu}\frac{l^{2}}{r^{4}}\left[\begin{array}{l}\left\{15-\left(15+15\dfrac{r}{l}+6\dfrac{r^{2}}{l^{2}}+\dfrac{r^{3}}{l^{3}}\right)e^{-r/l}\right\}\dfrac{x_{i}x_{j}x_{q}}{r^{3}}\\ +\left\{\left(3+3\dfrac{r}{l}+\dfrac{r^{2}}{l^{2}}\right)e^{-r/l}-3\right\}\dfrac{\delta_{ij}x_{q}+\delta_{jq}x_{i}+\delta_{iq}x_{j}}{r}\end{array}\right]a_{q} \quad (66)$$

$$+\frac{1}{8\pi\mu}\frac{1}{r^{2}}\left(1+\frac{r}{l}\right)e^{-r/l}\frac{x_{i}\delta_{jq}+x_{j}\delta_{iq}}{r}a_{q}$$

Now, it is easily seen that

$$e_{kk}^{F} = u_{k,k}^{F} = -\frac{1-2\nu}{8\pi\mu(1-\nu)}\frac{x_{q}}{r^{3}}a_{q} \quad (67)$$

Therefore, the symmetric part of force-stress tensor becomes

$$\sigma_{(ji)}^{F} = 2\mu\left(\frac{\nu}{1-2\nu}e_{kk}^{F}\delta_{ij}+e_{ij}^{F}\right)=$$

$$-\frac{1}{8\pi(1-\nu)}\frac{1}{r^{2}}\left[(1-2\nu)\frac{x_{j}\delta_{iq}+x_{i}\delta_{jq}-x_{q}\delta_{ij}}{r}+\frac{3x_{i}x_{j}x_{q}}{r^{3}}\right]a_{q}$$

$$+\frac{1}{2\pi}\frac{l^{2}}{r^{4}}\left[\begin{array}{l}\left\{15-\left(15+15\dfrac{r}{l}+6\dfrac{r^{2}}{l^{2}}+\dfrac{r^{3}}{l^{3}}\right)e^{-r/l}\right\}\dfrac{x_{i}x_{j}x_{q}}{r^{3}}\\ +\left\{\left(3+3\dfrac{r}{l}+\dfrac{r^{2}}{l^{2}}\right)e^{-r/l}-3\right\}\dfrac{\delta_{ij}x_{q}+\delta_{jq}x_{i}+\delta_{iq}x_{j}}{r}\end{array}\right]a_{q} \quad (68)$$

$$+\frac{1}{4\pi}\frac{1}{r^{2}}\left(1+\frac{r}{l}\right)e^{-r/l}\frac{x_{i}\delta_{jq}+x_{j}\delta_{iq}}{r}a_{q}$$

For the rotation vector, we have

$$\boldsymbol{\omega}^{F} = \frac{1}{2}\nabla\times\mathbf{u}^{F} = \frac{1}{2}\nabla\times\mathbf{u}^{(2)} = \frac{1}{2}\nabla\times\left(-\frac{1}{4\pi\mu}\frac{1-e^{-r/l}}{r}\mathbf{a}\right) \quad (69)$$

which gives

$$\omega_i^F = \frac{1}{8\pi\mu}\frac{1}{r^2}\left[\left(1+\frac{r}{l}\right)e^{-r/l}-1\right]\frac{\varepsilon_{ipq}x_p}{r}a_q \tag{70}$$

It is seen that the mean curvature vector is

$$\kappa_i^F = \frac{1}{2}\varepsilon_{ijk}\omega_{k,j}^F$$

$$= \frac{1}{16\pi\mu}\frac{1}{r^3}\left[3-\left(3+3\frac{r}{l}+\frac{r^2}{l^2}\right)e^{-r/l}\right]\frac{x_ix_q}{r^2}a_q + \frac{1}{16\pi\mu}\frac{1}{r^3}\left[\left(1+\frac{r}{l}+\frac{r^2}{l^2}\right)e^{-r/l}-1\right]\delta_{iq}a_q \tag{71}$$

and therefore, for the couple-stress vector, we have

$$\mu_i^F = -8\mu l^2\kappa_i^F$$

$$= \frac{1}{2\pi}\frac{l^2}{r^3}\left[\left(3+3\frac{r}{l}+\frac{r^2}{l^2}\right)e^{-r/l}-3\right]\frac{x_ix_q}{r^2}a_q - \frac{1}{2\pi}\frac{l^2}{r^3}\left[\left(1+\frac{r}{l}+\frac{r^2}{l^2}\right)e^{-r/l}-1\right]\delta_{iq}a_q \tag{72}$$

It is seen that the skew-symmetric part of the force-stress tensor becomes

$$\sigma_{[ji]}^F = -\mu_{[i,j]}^F = \frac{1}{4\pi}\frac{1}{r^2}\left[\left(1+\frac{r}{l}\right)e^{-r/l}\right]\frac{x_i\delta_{jq}-x_j\delta_{iq}}{r}a_q \tag{73}$$

Therefore, the total force-stress tensor is

$$\sigma_{ji}^F = \sigma_{(ji)}^F + \sigma_{[ji]}^F = -\frac{1}{8\pi(1-\nu)}\frac{1}{r^2}\left[(1-2\nu)\frac{x_j\delta_{iq}+x_i\delta_{jq}-x_q\delta_{ij}}{r}+\frac{3x_ix_jx_q}{r^3}\right]a_q$$

$$+\frac{1}{2\pi}\frac{l^2}{r^4}\left[\begin{array}{l}\left\{15-\left(15+15\dfrac{r}{l}+6\dfrac{r^2}{l^2}+\dfrac{r^3}{l^3}\right)e^{-r/l}\right\}\dfrac{x_ix_jx_q}{r^3}\\ +\left\{\left(3+3\dfrac{r}{l}+\dfrac{r^2}{l^2}\right)e^{-r/l}-3\right\}\dfrac{\delta_{ij}x_q+\delta_{jq}x_i+\delta_{iq}x_j}{r}\end{array}\right]a_q \tag{74}$$

$$+\frac{1}{2\pi}\frac{1}{r^2}\left[\left(1+\frac{r}{l}\right)e^{-r/l}\right]\frac{x_i\delta_{jq}}{r}a_q$$

Interestingly, the force-traction vector becomes

$$t_i^{(n)F} = \sigma_{ji}^F n_j =$$

$$-\frac{1}{8\pi(1-\nu)}\frac{1}{r^2}\left[(1-2\nu)\frac{x_j n_j \delta_{iq} + x_i n_q - x_q n_i}{r} + \frac{3x_i x_j x_q}{r^3}n_j\right]a_q$$

$$+\frac{1}{2\pi}\frac{l^2}{r^4}\left[\begin{array}{l}\left\{15-\left(15+15\dfrac{r}{l}+6\dfrac{r^2}{l^2}+\dfrac{r^3}{l^3}\right)e^{-r/l}\right\}\dfrac{x_i x_j x_q}{r^3}n_j \\ +\left\{\left(3+3\dfrac{r}{l}+\dfrac{r^2}{l^2}\right)e^{-r/l}-3\right\}\dfrac{n_i x_q + n_q x_i + \delta_{iq} x_j n_j}{r}\end{array}\right]a_q \quad (75)$$

$$+\frac{1}{2\pi}\frac{1}{r^2}\left(1+\frac{r}{l}\right)e^{-r/l}\frac{x_i n_q}{r}a_q$$

and the moment-traction vector is

$$m_i^{F(n)} = \varepsilon_{ijk} n_j \mu_k^F = \frac{1}{2\pi}\frac{l^2}{r^3}\left[3-\left(3+3\frac{r}{l}+\frac{r^2}{l^2}\right)e^{-r/l}\right]\left(\frac{x_j n_j}{r}\varepsilon_{ipq}-\frac{x_i n_j}{r}\varepsilon_{jpq}\right)\frac{x_p}{r}a_q$$

$$+\frac{1}{\pi}\frac{l^2}{r^3}\left[\left(1+\frac{r}{l}\right)e^{-r/l}-1\right]\varepsilon_{ijq}n_j a_q \quad (76)$$

Finally, we can consider the following relations

$$u_i^F = U_{iq}^F a_q \quad (77)$$

$$\omega_i^F = \Omega_{iq}^F a_q \quad (78)$$

$$\sigma_{ji}^F = \Sigma_{jiq}^F a_q \quad (79)$$

$$\mu_i^F = \mathsf{M}_{iq}^F a_q \quad (80)$$

$$t_i^{(n)F} = T_{iq}^F a_q \quad (81)$$

$$m_i^{(n)F} = M_{iq}^F a_q \quad (82)$$

where $U_{iq}^F$, $\Omega_{iq}^F$, $\Sigma_{jiq}^F$, $\mathsf{M}_{iq}^F$, $T_{iq}^F$ and $M_{iq}^F$ represent the corresponding displacement, rotation, force-stress, couple-stress, force-traction and moment-traction, respectively, at $x$ due to a unit concentrated force in the $q$-direction at the origin. It is seen that, these Green's functions are

$$U_{iq}^{F} = \frac{1}{16\pi\mu(1-\nu)}\frac{1}{r}\left[(3-4\nu)\delta_{iq} + \frac{x_i x_q}{r^2}\right]$$
$$+ \frac{1}{4\pi\mu}\frac{l^2}{r^3}\left[\left\{\left(3+3\frac{r}{l}+\frac{r^2}{l^2}\right)e^{-r/l}-3\right\}\frac{x_i x_q}{r^2} + \left\{1-\left(1+\frac{r}{l}+\frac{r^2}{l^2}\right)e^{-r/l}\right\}\delta_{iq}\right] \quad (83)$$

$$\Omega_{iq}^{F} = \frac{1}{8\pi\mu}\frac{1}{r^2}\left[\left(1+\frac{r}{l}\right)e^{-r/l}-1\right]\frac{\varepsilon_{ipq}x_p}{r} \quad (84)$$

$$\Sigma_{jiq}^{F} = -\frac{1}{8\pi(1-\nu)}\frac{1}{r^2}\left[(1-2\nu)\frac{x_j\delta_{iq}+x_i\delta_{jq}-x_q\delta_{ij}}{r}+\frac{3x_i x_j x_q}{r^3}\right]a_q$$
$$+ \frac{1}{2\pi}\frac{l^2}{r^4}\left[\begin{array}{l}\left\{15-\left(15+15\frac{r}{l}+6\frac{r^2}{l^2}+\frac{r^3}{l^3}\right)e^{-r/l}\right\}\frac{x_i x_j x_q}{r^3} \\ +\left\{\left(3+3\frac{r}{l}+\frac{r^2}{l^2}\right)e^{-r/l}-3\right\}\frac{\delta_{ij}x_q+\delta_{jq}x_i+\delta_{iq}x_j}{r}\end{array}\right]a_q \quad (85)$$
$$+ \frac{1}{2\pi}\frac{1}{r^2}\left[\left(1+\frac{r}{l}\right)e^{-r/l}\right]\frac{x_j\delta_{iq}}{r}a_q$$

$$\mathsf{M}_{iq}^{F} = \frac{1}{2\pi}\frac{l^2}{r^3}\left[\left(3+3\frac{r}{l}+\frac{r^2}{l^2}\right)e^{-r/l}-3\right]\frac{x_i x_q}{r^2} - \frac{1}{2\pi}\frac{l^2}{r^3}\left[\left(1+\frac{r}{l}+\frac{r^2}{l^2}\right)e^{-r/l}-1\right]\delta_{iq} \quad (86)$$

$$T_{iq}^{(n)F} = -\frac{1}{8\pi(1-\nu)}\frac{1}{r^2}\left[(1-2\nu)\frac{x_j n_j\delta_{iq}+x_i n_q-x_q n_i}{r}+\frac{3x_i x_j x_q}{r^3}n_j\right]$$
$$+ \frac{1}{2\pi}\frac{l^2}{r^4}\left[\begin{array}{l}\left\{15-\left(15+15\frac{r}{l}+6\frac{r^2}{l^2}+\frac{r^3}{l^3}\right)e^{-r/l}\right\}\frac{x_i x_q x_j n_j}{r^3} \\ +\left\{\left(3+3\frac{r}{l}+\frac{r^2}{l^2}\right)e^{-r/l}-3\right\}\frac{n_i x_q+n_q x_i+\delta_{iq}x_j n_j}{r}\end{array}\right] \quad (87)$$
$$+ \frac{1}{2\pi}\frac{1}{r^2}\left[\left(1+\frac{r}{l}\right)e^{-r/l}\right]\frac{x_i n_q}{r}$$

$$M_{iq}^{F} = \frac{1}{2\pi}\frac{l^2}{r^3}\left[3-\left(3+3\frac{r}{l}+\frac{r^2}{l^2}\right)e^{-r/l}\right]\left(\frac{x_j n_j}{r}\varepsilon_{ipq}-\frac{x_i n_j}{r}\varepsilon_{jpq}\right)\frac{x_p}{r}$$
$$+ \frac{1}{\pi}\frac{l^2}{r^3}\left[\left(1+\frac{r}{l}\right)e^{-r/l}-1\right]\varepsilon_{ijq}n_j \quad (88)$$

## 2.2 Concentrated Couple

Assume that in an infinitely extended couple stress elastic material there is a unit concentrated couple at the origin in the arbitrary direction specified by the unit vector **a**. Therefore, this concentrated couple can be represented as a body couple

$$\mathbf{C} = \delta^{(3)}(\mathbf{x})\mathbf{a} \tag{89}$$

As was mentioned previously, the effect of a body couple is equivalent to the effect of a body force represented by

$$\mathbf{F}^C = \frac{1}{2}\nabla \times \mathbf{C} = \frac{1}{2}\nabla \delta^{(3)}(\mathbf{x}) \times \mathbf{a} \tag{90}$$

or

$$F_i^C = \frac{1}{2}\varepsilon_{ijk}\delta^3(\mathbf{x})_{,j}a_k \tag{91}$$

with a vanishing surface effect at infinity. This shows that the displacement field of the concentrated couple $\mathbf{C} = \delta^{(3)}(\mathbf{x})\mathbf{a}$ is equivalent to the rotation field of the concentrated force $\mathbf{F} = \delta^{(3)}(\mathbf{x})\mathbf{a}$. Therefore, the solutions of the two problems are related, such that

$$u_i^C = \frac{1}{2}\varepsilon_{ijk}u_{k,j}^F = \omega_i^F \tag{92}$$

which gives

$$u_i^C = \frac{1}{8\pi\mu}\frac{1}{r^2}\left[\left(1+\frac{r}{l}\right)e^{-r/l} - 1\right]\frac{\varepsilon_{ipq}x_p}{r}a_q \tag{93}$$

From this it is seen that the gradient of the displacement is

$$u_{i,j}^C = \frac{1}{8\pi\mu}\frac{1}{r^3}\left[3-\left(3+3\frac{r}{l}+\frac{r^2}{l^2}\right)e^{-r/l}\right]\frac{x_p x_j}{r^2}\varepsilon_{ipq}a_q + \frac{1}{8\pi\mu}\frac{1}{r^3}\left[\left(1+\frac{r}{l}\right)e^{-r/l}-1\right]\varepsilon_{ijq}a_q \tag{94}$$

Therefore, the strain tensor becomes

$$e_{ij}^C = u_{(i,j)}^C = \frac{1}{16\pi\mu}\frac{1}{r^3}\left[3-\left(3+3\frac{r}{l}+\frac{r^2}{l^2}\right)e^{-r/l}\right]\left(\frac{x_j}{r}\varepsilon_{ipq}+\frac{x_i}{r}\varepsilon_{jpq}\right)\frac{x_p}{r}a_q \tag{95}$$

It is interesting to note that

$$e_{kk}^C = u_{k,k}^C = 0 \tag{96}$$

which means the deformation field of a concentrated couple $\mathbf{C} = \delta(\mathbf{x})\mathbf{a}$ is equivoluminal. Therefore, the symmetric part of force-stress tensor is

$$\sigma_{(ji)}^C = 2\mu e_{ij}^C = \frac{1}{8\pi}\frac{1}{r^3}\left[3-\left(3+3\frac{r}{l}+\frac{r^2}{l^2}\right)e^{-r/l}\right]\left(\frac{x_j x_p}{r^2}\varepsilon_{ipq} + \frac{x_i x_p}{r^2}\varepsilon_{jpq}\right)a_q \tag{97}$$

For the rotation vector, we have

$$\begin{aligned}\omega_i^C &= \frac{1}{2}\varepsilon_{ijk}u_{k,j}^C \\ &= \frac{1}{16\pi\mu}\frac{1}{r^3}\left[3-\left(3+3\frac{r}{l}+\frac{r^2}{l^2}\right)e^{-r/l}\right]\left(\frac{x_i x_q}{r^2}-\delta_{iq}\right)a_q - \frac{1}{8\pi\mu}\frac{1}{r^3}\left[\left(1+\frac{r}{l}\right)e^{-r/l}-1\right]\delta_{iq}a_q\end{aligned} \tag{98}$$

Furthermore, the mean curvature vector is

$$\kappa_i^C = \frac{1}{2}\varepsilon_{ijk}\omega_{k,j}^C = -\frac{1}{32\pi\mu l^2}\frac{1}{r^2}\left[\left(1+\frac{r}{l}\right)e^{-r/l}\right]\frac{\varepsilon_{ipq}x_p}{r}a_q \tag{99}$$

and therefore the couple-stress vector becomes

$$\mu_i^C = -8\mu l^2 \kappa_i^C = \frac{1}{4\pi}\frac{1}{r^2}\left[\left(1+\frac{r}{l}\right)e^{-r/l}\right]\frac{\varepsilon_{ipq}x_p}{r}a_q \tag{100}$$

The skew-symmetric part of $\mu_{i,j}^C$ is

$$\begin{aligned}\mu_{[i,j]}^C &= -\frac{1}{8\pi}\frac{1}{r^3}\left[\left(3+3\frac{r}{l}+\frac{r^2}{l^2}\right)e^{-r/l}\right]\left(\varepsilon_{ipq}\frac{x_j x_p}{r^2} - \varepsilon_{jpq}\frac{x_i x_p}{r^2}\right)a_q \\ &\quad + \frac{1}{4\pi}\frac{1}{r^3}\left[\left(1+\frac{r}{l}\right)e^{-r/l}\right]\varepsilon_{ijq}a_q\end{aligned} \tag{101}$$

Therefore, the skew-symmetric part of force-stress tensor becomes

$$\sigma_{[ji]}^{C} = -\mu_{[i,j]}^{C} = \frac{1}{8\pi} \frac{1}{r^3} \left[ \left( 3 + 3\frac{r}{l} + \frac{r^2}{l^2} \right) e^{-r/l} \right] \left( \varepsilon_{ipq} \frac{x_j x_p}{r^2} - \varepsilon_{jpq} \frac{x_i x_p}{r^2} \right) a_q$$
$$- \frac{1}{4\pi} \frac{1}{r^3} \left[ \left( 1 + \frac{r}{l} \right) e^{-r/l} \right] \varepsilon_{ijq} a_q \qquad (102)$$

and the total force-stress tensor becomes

$$\sigma_{ji}^{C} = \frac{1}{8\pi} \frac{3}{r^3} \frac{x_j x_p}{r^2} \varepsilon_{ipq} a_q + \frac{1}{8\pi} \frac{1}{r^3} \left[ 3 - 2 \left( 3 + 3\frac{r}{l} + \frac{r^2}{l^2} \right) e^{-r/l} \right] \frac{x_i x_p}{r^2} \varepsilon_{jpq} a_q$$
$$- \frac{1}{4\pi} \frac{1}{r^3} \left[ \left( 1 + \frac{r}{l} \right) e^{-r/l} \right] \varepsilon_{ijq} a_q \qquad (103)$$

Then, the force-traction vector becomes

$$t_i^{(C)} = \sigma_{ji}^{C} n_j$$
$$= \frac{1}{8\pi} \frac{3}{r^3} \frac{x_j x_p}{r^2} \varepsilon_{ipq} n_j a_q + \frac{1}{8\pi} \frac{1}{r^3} \left[ 3 - 2 \left( 3 + 3\frac{r}{l} + \frac{r^2}{l^2} \right) e^{-r/l} \right] \frac{x_i x_p}{r^2} \varepsilon_{jpq} n_j a_q \qquad (104)$$
$$- \frac{1}{4\pi} \frac{1}{r^3} \left[ \left( 1 + \frac{r}{l} \right) e^{-r/l} \right] \varepsilon_{ijq} n_j a_q$$

and the couple-traction vector is given by

$$m_i^{C(n)} = \varepsilon_{ijk} n_j \mu_k^{(C)} = \frac{1}{4\pi} \frac{1}{r^2} \left[ \left( 1 + \frac{r}{l} \right) e^{-r/l} \right] \frac{x_i \delta_{jq} - x_j \delta_{iq}}{r} n_j a_q \qquad (105)$$

Therefore, we can consider

$$u_i^{C} = U_{iq}^{C} a_q \qquad (106)$$

$$\omega_i^{C} = \Omega_{iq}^{C} a_q \qquad (107)$$

$$\sigma_{ji}^{C} = \Sigma_{jiq}^{C} a_q \qquad (108)$$

$$\mu_i^{C} = \mathsf{M}_{iq}^{C} a_q \qquad (109)$$

$$t_i^{(n)C} = T_{iq}^{C} a_q \qquad (110)$$

$$m_i^{(n)C} = M_{iq}^{C} a_q \qquad (111)$$

where $U_{iq}^C$, $\Omega_{iq}^C$, $\Sigma_{jiq}^C$, $\mathsf{M}_{iq}^C$, $T_{iq}^C$ and $M_{iq}^C$ represent the corresponding displacement, rotation, force-stress, couple-stress, force-traction and moment-traction, respectively, at $x$ due to a unit concentrated couple in the $q$ direction at the origin. Therefore, it is easily seen that these infinite space Green's functions are

$$U_{iq}^C = \Omega_{iq}^F = \frac{1}{8\pi\mu}\frac{1}{r^2}\left[\left(1+\frac{r}{l}\right)e^{-r/l} - 1\right]\varepsilon_{ipq}\frac{x_p}{r} \tag{112}$$

$$\Omega_{iq}^C = \frac{1}{16\pi\mu}\frac{1}{r^3}\left[3 - \left(3 + 3\frac{r}{l} + \frac{r^2}{l^2}\right)e^{-r/l}\right]\left(\frac{x_i x_q}{r^2} - \delta_{iq}\right) - \frac{1}{8\pi\mu}\frac{1}{r^3}\left[\left(1+\frac{r}{l}\right)e^{-r/l} - 1\right]\delta_{iq} \tag{113}$$

$$\begin{aligned}\Sigma_{jiq}^C &= \frac{1}{8\pi}\frac{3}{r^3}\frac{x_j x_p}{r^2}\varepsilon_{ipq} \\ &+ \frac{1}{8\pi}\frac{1}{r^3}\left[3 - 2\left(3 + 3\frac{r}{l} + \frac{r^2}{l^2}\right)e^{-r/l}\right]\frac{x_i x_p}{r^2}\varepsilon_{jpq} - \frac{1}{4\pi}\frac{1}{r^3}\left[\left(1+\frac{r}{l}\right)e^{-r/l}\right]\varepsilon_{ijq}\end{aligned} \tag{114}$$

$$\mathsf{M}_{iq}^C = \frac{1}{4\pi}\frac{1}{r^2}\left[\left(1+\frac{r}{l}\right)e^{-r/l}\right]\frac{\varepsilon_{ipq}x_p}{r} \tag{115}$$

$$\begin{aligned}T_{iq}^C &= \frac{1}{8\pi}\frac{3}{r^3}\frac{x_j x_p}{r^2}\varepsilon_{ipq}n_j \\ &+ \frac{1}{8\pi}\frac{1}{r^3}\left[3 - 2\left(3 + 3\frac{r}{l} + \frac{r^2}{l^2}\right)e^{-r/l}\right]\frac{x_i x_p}{r^2}\varepsilon_{jpq}n_j - \frac{1}{4\pi}\frac{1}{r^3}\left[\left(1+\frac{r}{l}\right)e^{-r/l}\right]\varepsilon_{ijq}n_j\end{aligned} \tag{116}$$

$$M_{iq}^C = \frac{1}{4\pi}\frac{1}{r^2}\left[\left(1+\frac{r}{l}\right)e^{-r/l}\right]\frac{x_i n_q - x_j n_j \delta_{iq}}{r} \tag{117}$$

## 3. Fundamental Solutions for Two-Dimensional Case

In this section, we first present the governing equations of size-dependent couple stress elasticity in two-dimensions under plane strain conditions. Then, we derive the complete two-dimensional fundamental solutions in a similar method used in the three-dimensional case. Reference [19] uses these fundamental solutions within a two-dimensional boundary element method.

## 3.1 Governing Equations for Two Dimensions

We suppose that the media occupies a cylindrical region, such that the axis of the cylinder is parallel to the $x_3$-axis. Furthermore, we assume the body is in a state of planar deformation parallel to this plane, such that

$$u_{\alpha,3} = 0, \ u_3 = 0 \quad \text{in} \ V \tag{118a,b}$$

where all Greek indices, here and throughout the remainder of the paper, will vary only over (1,2). Also, let $V^{(2)}$ and $S^{(2)}$ represent, respectively, the cross section of the body in the $x_1 x_2$-plane and its bounding edge in that plane.

As a result of these assumptions, all quantities are independent of $x_3$. Then, throughout the domain

$$\omega_\alpha = 0, \ e_{3i} = e_{i3} = 0, \ \kappa_3 = 0 \tag{119a-c}$$

and

$$\sigma_{3\alpha} = \sigma_{\alpha 3} = 0, \ \mu_3 = \mu_{21} = 0 \tag{120a,b}$$

Introducing the abridged notation

$$\omega = \omega_3 = \frac{1}{2}\varepsilon_{\alpha\beta} u_{\beta,\alpha} \tag{121}$$

where $\varepsilon_{\alpha\beta}$ is the two-dimensional alternating symbol with

$$\varepsilon_{12} = -\varepsilon_{21} = 1, \quad \varepsilon_{11} = \varepsilon_{22} = 0 \tag{122}$$

Now, it is seen that the non-zero components of the curvature vector are

$$\kappa_\alpha = \frac{1}{2}\varepsilon_{\alpha\beta}\omega_{,\beta} \tag{123}$$

Therefore, the non-zero components of stresses are written

$$\mu_\alpha = -4\eta\varepsilon_{\alpha\beta}\omega_{,\beta} \tag{124}$$

$$\sigma_{(\beta\alpha)} = \lambda e_{\gamma\gamma}\delta_{\alpha\beta} + 2\mu e_{\alpha\beta} \tag{125}$$

$$\sigma_{[\beta\varepsilon]} = -\mu_{[\alpha,\beta]} = 2\eta\varepsilon_{\alpha\beta}\nabla^2\omega \tag{126}$$

and

$$\sigma_{\beta\alpha} = \lambda e_{\gamma\gamma}\delta_{\alpha\beta} + 2\mu e_{\alpha\beta} - 2\eta\varepsilon_{\beta\alpha}\nabla^2\omega \tag{127}$$

All the other stresses are zero, apart from $\sigma_{33}$ and $\mu_{3\alpha}$, which are given as

$$\sigma_{33} = \nu\sigma_{\gamma\gamma} \tag{128}$$

$$\mu_{3\alpha} = -4\eta\omega_{,\alpha} \tag{129}$$

It should be noticed that these stresses in Eqs. (128) and (129), acting on planes parallel to the $x_1x_2$-plane, do not enter directly into the solution of the boundary value problem.

For the planar problem, the stresses must satisfy the three equilibrium equations

$$\sigma_{\beta\alpha,\beta} + F_\alpha = 0 \tag{130}$$

$$\mu_{\beta 3,\beta} + \varepsilon_{\alpha\beta}\sigma_{\alpha\beta} = 0 \tag{131}$$

with the obvious requirement $F_3 = 0$. The moment equation can be written as

$$\sigma_{[\beta\alpha]} = -\mu_{[\alpha,\beta]} \tag{132}$$

which actually gives the non-zero components

$$\sigma_{[21]} = -\sigma_{[12]} = -\mu_{[1,2]} \tag{133}$$

We also notice that the force-traction reduces to

$$t_\alpha^{(n)} = \sigma_{\beta\alpha}n_\beta \tag{134}$$

and the moment-traction has only one component $m_3$. This can be conveniently denoted by the abridged symbol $m$, where

$$m^{(n)} = m_3^{(n)} = \varepsilon_{\beta\alpha}\mu_\alpha n_\beta = 4\eta\frac{\partial\omega}{\partial n} \tag{135}$$

*3.2 Concentrated Force*

Assume that there is a line load on the $x_3$ axis with an intensity of unity per unit length in the arbitrary direction specified by the unit vector **a**. This distributed load can be represented by the body force

$$\mathbf{F} = \delta^{(2)}(\mathbf{x})\mathbf{a} \tag{136}$$

where $\delta^{(2)}(\mathbf{x})$ is the Dirac delta function in two-dimensional space. It is known that

$$\nabla^2\left(\frac{1}{2\pi}\ln r\right) = \delta^{(2)}(\mathbf{x}) \tag{137}$$

By applying the vectorial relation in Eq. (39) for the vector $\frac{1}{2\pi}\ln r\,\mathbf{a}$, we decompose the body force as

$$\mathbf{F} = \nabla^2\left(\frac{1}{2\pi}\ln r\right)\mathbf{a} = \nabla\left(\nabla\bullet\frac{\ln r\,\mathbf{a}}{2\pi}\right) - \nabla\times\left(\nabla\times\frac{\ln r\,\mathbf{a}}{2\pi}\right) \tag{138}$$

If we consider the decomposition of displacement $\mathbf{u}^F$ as

$$\mathbf{u}^F = \mathbf{u}^{(1)} + \mathbf{u}^{(2)} \tag{139}$$

where $\mathbf{u}^{(1)}$ and $\mathbf{u}^{(2)}$ are the dilatational and solenoidal part of the displacement vector $\mathbf{u}^F$ satisfying

$$\nabla\times\mathbf{u}^{(1)} = 0 \tag{140}$$

$$\nabla\bullet\mathbf{u}^{(2)} = 0 \tag{141}$$

then it is seen that

$$(\lambda + 2\mu)\nabla^2\mathbf{u}^{(1)} = -\nabla\left(\nabla\bullet\frac{\ln r\,\mathbf{a}}{2\pi}\right) \tag{142}$$

$$\eta\nabla^2\nabla^2\mathbf{u}^{(2)} - \mu\nabla^2\mathbf{u}^{(2)} = -\nabla\times\left(\nabla\times\frac{\ln r\,\mathbf{a}}{2\pi}\right) \tag{143}$$

If we introduce two vectors $\mathbf{A}^{(1)}$ and $\mathbf{A}^{(2)}$ such that

$$\mathbf{u}^{(1)} = \nabla\left(\nabla\bullet\mathbf{A}^{(1)}\right) \tag{144}$$

$$\begin{aligned}\mathbf{u}^{(2)} &= \nabla\times\left(\nabla\times\mathbf{A}^{(2)}\right) \\ &= \nabla\left(\nabla\bullet\mathbf{A}^{(2)}\right) - \nabla^2\mathbf{A}^{(2)}\end{aligned} \tag{145}$$

which satisfy the conditions of Eqs. (140) and (141), we obtain

$$(\lambda + 2\mu)\nabla^2\mathbf{A}^{(1)} = -\frac{\ln r}{2\pi}\mathbf{a} \tag{146}$$

$$\eta\nabla^2\nabla^2\mathbf{A}^{(2)} - \mu\nabla^2\mathbf{A}^{(2)} = -\frac{\ln r}{2\pi}\mathbf{a} \tag{147}$$

Then, the solutions are in the form

$$\mathbf{A}^{(1)} = \varphi \mathbf{a} \tag{148}$$

$$\mathbf{A}^{(2)} = \psi \mathbf{a} \tag{149}$$

where $\varphi$ and $\psi$ are scalar functions of $r$ having two-dimensional radial symmetry. Therefore

$$\nabla^2 \varphi = -\frac{1}{2\pi(\lambda + 2\mu)} \ln r \tag{150}$$

$$\eta \nabla^2 \nabla^2 \psi - \mu \nabla^2 \psi = -\frac{\ln r}{2\pi} \tag{151}$$

In two-dimensions, with radial symmetry, the laplacian $\nabla^2$ reduces to

$$\nabla^2 \to \frac{d^2}{\partial r^2} + \frac{1}{r}\frac{d}{dr} \tag{152}$$

The regular solutions to Eqs. (150) and (151) are

$$\varphi = -\frac{1}{8\pi(\lambda + 2\mu)}\left(r^2 \ln r - r^2\right) \tag{153}$$

$$\psi = \frac{l^2}{2\pi\mu}\left[K_0\left(\frac{r}{l}\right) + \ln r\right] + \frac{1}{8\pi\mu}\left(r^2 \ln r - r^2\right) \tag{154}$$

where $K_0$ is the modified Bessel function of zeroth order. Therefore,

$$\mathbf{A}^{(1)} = -\frac{1}{8\pi(\lambda + 2\mu)}\left(r^2 \ln r - r^2\right)\mathbf{a} \tag{155}$$

$$\mathbf{A}^{(2)} = \frac{l^2}{2\pi\mu}\left[K_0\left(\frac{r}{l}\right) + \ln r\right]\mathbf{a} + \frac{1}{8\pi\mu}\left(r^2 \ln r - r^2\right)\mathbf{a} \tag{156}$$

Then, from Eq. (144)

$$u_\alpha^{(1)} = -\frac{1-2\nu}{16\pi\mu(1-\nu)}\left[2\frac{x_\alpha x_\rho}{r^2} + (2\ln r - 1)\delta_{\alpha\rho}\right]a_\rho \tag{157}$$

and from Eq. (145)

$$\mathbf{u}^{(2)} = \nabla\left(\nabla \bullet \mathbf{A}^{(2)}\right) - \frac{1}{2\pi\mu}\left[K_0\left(\frac{r}{l}\right) + \ln r\right]\mathbf{a} \tag{158}$$

By using the relations

$$\frac{\partial}{\partial r} K_0\left(\frac{r}{l}\right) = -\frac{1}{l} K_1\left(\frac{r}{l}\right) \tag{159}$$

$$\frac{\partial}{\partial r} K_1\left(\frac{r}{l}\right) = -\frac{1}{l} K_0\left(\frac{r}{l}\right) - \frac{1}{r} K_1\left(\frac{r}{l}\right) \tag{160}$$

where $K_1$ is the modified Bessel function of first order, we obtain

$$u_\alpha^{(2)} = +\frac{1}{8\pi\mu}\left[2\frac{x_\alpha x_\rho}{r^2} - (2\ln r + 1)\delta_{\alpha\rho}\right]a_\rho$$
$$+ \frac{1}{2\pi\mu}\left[K_0\left(\frac{r}{l}\right) + \frac{2l}{r}K_1\left(\frac{r}{l}\right) - \frac{2l^2}{r^2}\right]\frac{x_\alpha x_\rho}{r^2}a_\rho - \frac{1}{2\pi\mu}\left[K_0\left(\frac{r}{l}\right) + \frac{l}{r}K_1\left(\frac{r}{l}\right) - \frac{l^2}{r^2}\right]\delta_{\alpha\rho}a_\rho \tag{161}$$

It should be noticed that there are rigid body translation terms in $u_\alpha^{(1)}$ and $u_\alpha^{(2)}$ which cannot affect stress distrributions. These terms can be neglected in this Green's function for stress analysis. Therefore, by ignoring these rigid body terms and using

$$u_\alpha^F = u_\alpha^{(1)} + u_\alpha^{(2)} \tag{162}$$

we obtain

$$u_\alpha^F = -\frac{1}{8\pi\mu(1-v)}\left[(3-4v)\ln r\,\delta_{\alpha\rho} - \frac{x_\alpha x_\rho}{r^2}\right]a_\rho$$
$$+ \frac{1}{2\pi\mu}\left[K_0\left(\frac{r}{l}\right) + \frac{2l}{r}K_1\left(\frac{r}{l}\right) - \frac{2l^2}{r^2}\right]\frac{x_\alpha x_\rho}{r^2}a_\rho - \frac{1}{2\pi\mu}\left[K_0\left(\frac{r}{l}\right) + \frac{l}{r}K_1\left(\frac{r}{l}\right) - \frac{l^2}{r^2}\right]\delta_{\alpha\rho}a_\rho \tag{163}$$

For the gradient of displacements, we have

$$u_{\alpha,\beta}^F = -\frac{1}{8\pi\mu(1-v)r}\left[(3-4v)\frac{x_\beta\delta_{\alpha\rho}}{r} - \frac{x_\alpha\delta_{\beta\rho} + x_\rho\delta_{\alpha\beta}}{r} + 2\frac{x_\alpha x_\beta x_\rho}{r^3}\right]a_q$$
$$+ \frac{1}{2\pi\mu r}\left[K_0\left(\frac{r}{l}\right) + \frac{2l}{r}K_1\left(\frac{r}{l}\right) - \frac{2l^2}{r^2}\right]\left(\frac{\delta_{ij}x_q + \delta_{jq}x_i + x_j\delta_{iq}}{r} - 4\frac{x_\alpha x_\beta x_\rho}{r^3}\right)a_q \tag{164}$$
$$+ \frac{1}{2\pi\mu l}K_1\left(\frac{r}{l}\right)\left(\frac{x_\beta\delta_{\alpha\rho}}{r} - \frac{x_\alpha x_\beta x_\rho}{r^3}\right)a_q$$

Therefore, the strain tensor becomes

$$u_{(\alpha,\beta)}^F = e_{\alpha\beta}^F = -\frac{1}{8\pi\mu(1-\nu)r}\left[(1-2\nu)\frac{x_\beta\delta_{\alpha\rho}+x_\alpha\delta_{\beta\rho}}{r} - \frac{\delta_{\alpha\beta}x_\rho}{r} + 2\frac{x_\alpha x_\beta x_\rho}{r^3}\right]a_\rho$$

$$+\frac{1}{2\pi\mu r}\left[K_0\left(\frac{r}{l}\right)+\frac{2l}{r}K_1\left(\frac{r}{l}\right)-\frac{2l^2}{r^2}\right]\left(\frac{\delta_{\alpha\beta}x_\rho+\delta_{\beta\rho}x_\alpha+\delta_{\alpha\rho}x_\beta}{r} - 4\frac{x_\alpha x_\beta x_\rho}{r^3}\right)a_\rho \quad (165)$$

$$+\frac{1}{4\pi\mu l}K_1\left(\frac{r}{l}\right)\left(\frac{\delta_{\beta\rho}x_\alpha+\delta_{\alpha\rho}x_\beta}{r} - \frac{2x_\alpha x_\beta x_\rho}{r^3}\right)a_\rho$$

This relation shows that

$$e_{\gamma\gamma}^{(F)} = e_{\gamma\gamma}^{(1)} = -\frac{1-2\nu}{4\pi\mu(1-\nu)}\frac{x_\rho}{r^2}a_\rho \quad (166)$$

Therefore, the symmetric part of the force-stress tensor is

$$\sigma_{(\beta\alpha)}^F = 2\mu\left(\frac{\nu}{1-2\nu}e_{\gamma\gamma}^F\delta_{\alpha\beta}+e_{\alpha\beta}^F\right) =$$

$$-\frac{1}{4\pi(1-\nu)r}\left[(1-2\nu)\frac{\delta_{\beta\rho}x_\alpha+\delta_{\alpha\rho}x_\beta-\delta_{\alpha\beta}x_\rho}{r} + 2\frac{x_\alpha x_\beta x_\rho}{r^3}\right]a_\rho$$

$$+\frac{1}{\pi r}\left[K_0\left(\frac{r}{l}\right)+\frac{2l}{r}K_1\left(\frac{r}{l}\right)-\frac{2l^2}{r^2}\right]\left(\frac{\delta_{\alpha\beta}x_\rho+\delta_{\beta\rho}x_\alpha+\delta_{\alpha\rho}x_\beta}{r} - 4\frac{x_\alpha x_\beta x_\rho}{r^3}\right)a_q \quad (167)$$

$$+\frac{1}{2\pi l}K_1\left(\frac{r}{l}\right)\left(\frac{\delta_{\beta\rho}x_\alpha+\delta_{\alpha\rho}x_\beta}{r} - 2\frac{x_\alpha x_\beta x_\rho}{r^3}\right)a_q$$

Next, we consider rotations and note that the only non-zero in-plane component is

$$\omega^F = \frac{1}{2}\varepsilon_{\alpha\beta}u_{\beta,\alpha}^F = \frac{1}{4\pi\mu l}\left[K_1\left(\frac{r}{l}\right)-\frac{l}{r}\right]\frac{\varepsilon_{\alpha\rho}x_\alpha}{r}a_\rho \quad (168)$$

Then, the mean curvature components are

$$\kappa_\alpha^F = \frac{1}{2}\varepsilon_{\alpha\beta}\omega_{,\beta}^F = -\frac{1}{8\pi\mu l^2}\left[K_0\left(\frac{r}{l}\right)+\frac{2l}{r}K_1\left(\frac{r}{l}\right)-\frac{2l^2}{r^2}\right]\frac{x_\alpha x_\rho}{r^2}a_\rho$$

$$+\frac{1}{8\pi\mu l^2}\left[K_0\left(\frac{r}{l}\right)+\frac{l}{r}K_1\left(\frac{r}{l}\right)-\frac{l^2}{r^2}\right]\delta_{\alpha\rho}a_\rho \quad (169)$$

Therefore, the couple-stress vector becomes

$$\mu_\alpha^F = -8\mu l^2 \kappa_\alpha^F$$

$$= \frac{1}{\pi}\left[K_0\left(\frac{r}{l}\right) + \frac{2l}{r}K_1\left(\frac{r}{l}\right) - \frac{2l^2}{r^2}\right]\frac{x_\alpha x_\rho}{r^2}a_\rho - \frac{1}{\pi}\left[K_0\left(\frac{r}{l}\right) + \frac{l}{r}K_1\left(\frac{r}{l}\right) - \frac{l^2}{r^2}\right]\delta_{\alpha\rho}a_\rho \quad (170)$$

and the skew-symmetric part of the force-stress tensor becomes

$$\sigma_{[\beta\alpha]}^F = -\mu_{[\alpha,\beta]}^F = \frac{1}{2\pi l}K_1\left(\frac{r}{l}\right)\frac{x_\alpha\delta_{\beta\rho} - x_\beta\delta_{\alpha\rho}}{r}a_\rho \quad (171)$$

with non-zero components

$$\sigma_{[21]}^F = -\sigma_{[12]}^F = \frac{1}{2\pi l}K_1\left(\frac{r}{l}\right)\frac{x_1 a_2 - x_2 a_1}{r} \quad (172)$$

Therefore, the total force-stress tensor is

$$\sigma_{\beta\alpha}^F = \sigma_{(\beta\alpha)}^F + \sigma_{[\beta\alpha]}^F =$$

$$-\frac{1}{4\pi(1-\nu)r}\left[(1-2\nu)\frac{\delta_{\beta\rho}x_\alpha + \delta_{\alpha\rho}x_\beta - \delta_{\alpha\beta}x_\rho}{r} + 2\frac{x_\alpha x_\beta x_\rho}{r^3}\right]a_\rho$$

$$+\frac{1}{\pi r}\left[K_0\left(\frac{r}{l}\right) + \frac{2l}{r}K_1\left(\frac{r}{l}\right) - \frac{2l^2}{r^2}\right]\left(\frac{\delta_{\alpha\beta}x_\rho + \delta_{\beta\rho}x_\alpha + \delta_{\alpha\rho}x_\beta}{r} - \frac{4x_\alpha x_\beta x_\rho}{r^3}\right)a_q \quad (173)$$

$$+\frac{1}{\pi l}K_1\left(\frac{r}{l}\right)\left(\frac{\delta_{\beta\rho}x_\alpha}{r} - \frac{x_\alpha x_\beta x_\rho}{r^3}\right)a_q$$

and for the force-traction vector, we have

$$t_\alpha^{(n)F} = \sigma_{\beta\alpha}^F n_\beta =$$

$$-\frac{1}{4\pi(1-\nu)r}\left[(1-2\nu)\frac{n_\rho x_\alpha + \delta_{\alpha\rho}x_\beta n_\beta - n_\alpha x_\rho}{r} + 2\frac{x_\alpha x_\rho x_\beta n_\beta}{r^3}\right]a_\rho$$

$$+\frac{1}{\pi r}\left[K_0\left(\frac{r}{l}\right) + \frac{2l}{r}K_1\left(\frac{r}{l}\right) - \frac{2l^2}{r^2}\right]\left(\frac{n_\alpha x_\rho + n_\rho x_\alpha + \delta_{\alpha\rho}x_\beta n_\beta}{r} - \frac{4x_\alpha x_\rho x_\beta n_\beta}{r^3}\right)a_\rho \quad (174)$$

$$+\frac{1}{\pi l}K_1\left(\frac{r}{l}\right)\left(\frac{n_\rho x_\alpha}{r} - \frac{x_\alpha x_\rho x_\beta n_\beta}{r^3}\right)a_q$$

while for moment-traction, we obtain

$$m^{(n)} = \varepsilon_{\beta\alpha}\mu_\alpha^F n_\beta = \frac{1}{\pi}\left[K_0\left(\frac{r}{l}\right)+\frac{2l}{r}K_1\left(\frac{r}{l}\right)-\frac{2l^2}{r^2}\right]\frac{\varepsilon_{\beta\alpha}x_\alpha x_\rho n_\beta}{r^2}a_\rho$$
$$-\frac{1}{\pi}\left[K_0\left(\frac{r}{l}\right)+\frac{l}{r}K_1\left(\frac{r}{l}\right)-\frac{l^2}{r^2}\right]\varepsilon_{\beta\rho}n_\beta a_\rho \quad (175)$$

Finally we can consider the following relations

$$u_\alpha^F = U_{\alpha\rho}^F a_\rho \quad (176)$$

$$\omega^F = \Omega_\rho^F a_\rho \quad (177)$$

$$\sigma_{\beta\alpha}^F = \Sigma_{\beta\alpha\rho}^F a_\rho \quad (178)$$

$$\mu_\alpha^F = \mathsf{M}_{\alpha\rho}^F a_\rho \quad (179)$$

$$t_\alpha^{(n)F} = T_{\alpha\rho}^F a_\rho \quad (180)$$

$$m^{(n)F} = M_\rho^F a_\rho \quad (181)$$

where the fundamental solutions $U_{\alpha\rho}^F$, $\Sigma_{\beta\alpha\rho}^F$, $\mathsf{M}_{\alpha\rho}^F$ and $T_{\alpha\rho}^F$ represent the corresponding displacement, force-stress, couple-stress and force-traction, respectively, at $x$ due to a unit concentrated force in the $\rho$-direction at the origin. Furthermore, the Green's functions $\Omega_\rho^F$ and $M_\rho^F$ represent the corresponding rotation and moment-traction, respectively, at $x$. From the above relations, one can establish

$$U_{\alpha\rho}^F = -\frac{1}{8\pi\mu(1-\nu)}\left[(3-4\nu)\ln r\delta_{\alpha\rho}-\frac{x_\alpha x_\rho}{r^2}\right]$$
$$+\frac{1}{2\pi\mu}\left[K_0\left(\frac{r}{l}\right)+\frac{2l}{r}K_1\left(\frac{r}{l}\right)-\frac{2l^2}{r^2}\right]\frac{x_\alpha x_\rho}{r^2}-\frac{1}{2\pi\mu}\left[K_0\left(\frac{r}{l}\right)+\frac{l}{r}K_1\left(\frac{r}{l}\right)-\frac{l^2}{r^2}\right]\delta_{\alpha\rho} \quad (182)$$

$$\Omega_\rho^F = \frac{1}{4\pi\mu l}\left[K_1\left(\frac{r}{l}\right)-\frac{l}{r}\right]\frac{\varepsilon_{\alpha\rho}x_\alpha}{r} \quad (183)$$

$$\Sigma_{\beta\alpha\rho}^{F} = -\frac{1}{4\pi(1-\nu)r}\left[(1-2\nu)\frac{\delta_{\beta\rho}x_\alpha + \delta_{\alpha\rho}x_\beta - \delta_{\alpha\beta}x_\rho}{r} + 2\frac{x_\alpha x_\beta x_\rho}{r^3}\right]$$

$$+ \frac{1}{\pi r}\left[K_0\left(\frac{r}{l}\right) + \frac{2l}{r}K_1\left(\frac{r}{l}\right) - \frac{2l^2}{r^2}\right]\left(\frac{\delta_{\alpha\beta}x_\rho + \delta_{\beta\rho}x_\alpha + \delta_{\alpha\rho}x_\beta}{r} - \frac{4x_\alpha x_\beta x_\rho}{r^3}\right) \quad (184)$$

$$+ \frac{1}{\pi l}K_1\left(\frac{r}{l}\right)\left(\frac{\delta_{\beta\rho}x_\alpha}{r} - \frac{x_\alpha x_\beta x_\rho}{r^3}\right)$$

$$\mathsf{M}_{\alpha\rho}^{F} = \frac{1}{\pi}\left[K_0\left(\frac{r}{l}\right) + \frac{2l}{r}K_1\left(\frac{r}{l}\right) - \frac{2l^2}{r^2}\right]\frac{x_\alpha x_\rho}{r^2} - \frac{1}{\pi}\left[K_0\left(\frac{r}{l}\right) + \frac{l}{r}K_1\left(\frac{r}{l}\right) - \frac{l^2}{r^2}\right]\delta_{\alpha\rho} \quad (185)$$

$$T_{\alpha\rho}^{F} = -\frac{1}{4\pi(1-\nu)r}\left[(1-2\nu)\frac{n_\rho x_\alpha + \delta_{\alpha\rho}x_\beta n_\beta - n_\alpha x_\rho}{r} + 2\frac{x_\alpha x_\rho x_\beta n_\beta}{r^3}\right]$$

$$+ \frac{1}{\pi r}\left[K_0\left(\frac{r}{l}\right) + \frac{2l}{r}K_1\left(\frac{r}{l}\right) - \frac{2l^2}{r^2}\right]\left(\frac{n_\alpha x_\rho + n_\rho x_\alpha + \delta_{\alpha\rho}x_\beta n_\beta}{r} - \frac{4x_\alpha x_\rho x_\beta n_\beta}{r^3}\right) \quad (186)$$

$$+ \frac{1}{\pi l}K_1\left(\frac{r}{l}\right)\left(\frac{n_\rho x_\alpha}{r} - \frac{x_\alpha x_\rho x_\beta n_\beta}{r^3}\right)$$

$$M_\rho^F = \frac{1}{\pi}\left[K_0\left(\frac{r}{l}\right) + \frac{2l}{r}K_1\left(\frac{r}{l}\right) - \frac{2l^2}{r^2}\right]\frac{\varepsilon_{\beta\alpha}x_\alpha x_\rho n_\beta}{r^2}$$

$$- \frac{1}{\pi}\left[K_0\left(\frac{r}{l}\right) + \frac{l}{r}K_1\left(\frac{r}{l}\right) - \frac{l^2}{r^2}\right]\varepsilon_{\beta\rho}n_\beta \quad (187)$$

### *3.3 Concentrated Couple*

Now, assume that there is a line distribution of couple load along the $x_3$ axis with unit intensity per unit length. This distributed load can be represented by a body couple

$$\mathbf{C} = \delta^{(2)}(\mathbf{x})\mathbf{e}_3 \quad (188)$$

As we mentioned earlier, similar to three-dimensional case, the effect of a body couple in an infinite domain is equivalent to the result of a body force represented by

$$\mathbf{F}^C = \frac{1}{2}\nabla\times\mathbf{C} = \frac{1}{2}\nabla\delta^{(2)}(\mathbf{x})\times\mathbf{e}_3 = \frac{1}{2}\varepsilon_{\alpha\beta}\delta^{(2)}_{,\beta}\mathbf{e}_\alpha \quad (189)$$

Therefore, it is seen that the solutions of the two problems of concentrated force $\mathbf{F} = \delta^{(2)}(\mathbf{x})\mathbf{a}$ and concentrated couple $\mathbf{C} = \delta^{(2)}(\mathbf{x})\mathbf{e}_3$ are related, such that

$$\mathbf{u}^C = \frac{1}{2}\varepsilon_{\alpha\beta}U^F_{\gamma\alpha,\beta}\mathbf{e}_\gamma \tag{190}$$

or

$$u^C_\alpha = \frac{1}{2}\varepsilon_{\gamma\beta}U^F_{\alpha\gamma,\beta} \tag{191}$$

Thus, we have

$$u^C_\alpha = \frac{1}{4\pi\mu l}\left[K_1\left(\frac{r}{l}\right) - \frac{l}{r}\right]\frac{\varepsilon_{\alpha\gamma}x_\gamma}{r} \tag{192}$$

and we can write the gradient of displacement as

$$u^C_{\alpha,\beta} = -\frac{1}{4\pi\mu l^2}\left[K_0\left(\frac{r}{l}\right) + \frac{2l}{r}K_1\left(\frac{r}{l}\right) - \frac{2l^2}{r^2}\right]\frac{\varepsilon_{\alpha\gamma}x_\gamma x_\beta}{r^2} + \frac{1}{4\pi\mu l}\frac{1}{r}\left[K_1\left(\frac{r}{l}\right) - \frac{l}{r}\right]\varepsilon_{\alpha\beta} \tag{193}$$

Therefore, the strain tensor is

$$e^C_{\alpha\beta} = -\frac{1}{8\pi\mu l^2}\left[K_0\left(\frac{r}{l}\right) + \frac{2l}{r}K_1\left(\frac{r}{l}\right) - \frac{2l^2}{r^2}\right]\frac{\varepsilon_{\alpha\gamma}x_\gamma x_\beta + \varepsilon_{\beta\gamma}x_\gamma x_\alpha}{r^2} \tag{194}$$

It is interesting to note that

$$e^C_{\gamma\gamma} = u^C_{\gamma,\gamma} = 0 \tag{195}$$

which means the deformation field of a concentrated couple is equivoluminal, a property shared by the three-dimensional case. Then, the symmetric part of the force-stress tensor is

$$\sigma^C_{(\beta\alpha)} = 2\mu e^C_{\beta\alpha} = -\frac{1}{4\pi l^2}\left[K_0\left(\frac{r}{l}\right) + \frac{2l}{r}K_1\left(\frac{r}{l}\right) - \frac{2l^2}{r^2}\right]\frac{\varepsilon_{\alpha\gamma}x_\gamma x_\beta + \varepsilon_{\beta\gamma}x_\gamma x_\alpha}{r^2} \tag{196}$$

We also have for in-plane rotation

$$\omega^C = \frac{1}{2}\varepsilon_{\alpha\beta}u^C_{\beta,\alpha} = \frac{1}{8\pi\mu l^2}K_0\left(\frac{r}{l}\right) \tag{197}$$

and the mean curvature vector is

$$\kappa_\alpha^C = \frac{1}{2}\varepsilon_{\alpha\beta}\omega_{,\beta}^C = -\frac{1}{16\pi\mu l^3}K_1\left(\frac{r}{l}\right)\frac{\varepsilon_{\alpha\beta}x_\beta}{r} \tag{198}$$

Therefore, the couple-stress vector becomes

$$\mu_\alpha^C = -8\mu l^2 \kappa_\alpha^C = \frac{1}{2\pi l}K_1\left(\frac{r}{l}\right)\frac{\varepsilon_{\alpha\beta}x_\beta}{r} \tag{199}$$

For the skew-symmetric part of force-stress tensor, we have

$$\sigma_{[\beta\alpha]}^F = -\mu_{[\alpha,\beta]}^F = \frac{1}{4\pi l^2}K_0\left(\frac{r}{l}\right)\varepsilon_{\alpha\beta} \tag{200}$$

with the non-zero components

$$\sigma_{[21]}^C = -\sigma_{[12]}^C = \frac{1}{4\pi l^2}K_0\left(\frac{r}{l}\right) \tag{201}$$

Therefore, for the total force-stress tensor, we have

$$\begin{aligned}\sigma_{\beta\alpha}^C &= \sigma_{(\beta\alpha)}^C + \sigma_{[\beta\alpha]}^C \\ &= -\frac{1}{4\pi l^2}\left[K_0\left(\frac{r}{l}\right) + \frac{2l}{r}K_1\left(\frac{r}{l}\right) - \frac{2l^2}{r^2}\right]\frac{\varepsilon_{\alpha\gamma}x_\gamma x_\beta + \varepsilon_{\beta\gamma}x_\gamma x_\alpha}{r^2} + \frac{1}{4\pi l^2}K_0\left(\frac{r}{l}\right)\varepsilon_{\alpha\beta}\end{aligned} \tag{202}$$

The force-traction vector then becomes

$$\begin{aligned}t_\alpha^{(n)C} &= T_\alpha^C = \sigma_{\beta\alpha}^C n_\beta \\ &= -\frac{1}{4\pi l^2}\left[K_0\left(\frac{r}{l}\right) + \frac{2l}{r}K_1\left(\frac{r}{l}\right) - \frac{2l^2}{r^2}\right]\frac{\varepsilon_{\alpha\gamma}x_\gamma x_\beta n_\beta + \varepsilon_{\beta\gamma}x_\gamma x_\alpha n_\beta}{r^2} + \frac{1}{4\pi l^2}K_0\left(\frac{r}{l}\right)\varepsilon_{\alpha\beta}n_\beta\end{aligned} \tag{203}$$

and the moment-traction is given by

$$m^{(n)C} = \varepsilon_{\beta\alpha}\mu_\alpha^F n_\beta = -\frac{1}{2\pi l}K_1\left(\frac{r}{l}\right)\frac{x_\beta n_\beta}{r} \tag{204}$$

Finally, we have all of the necessary Green's functions or influence functions for the two-dimensional case, which can be written

$$U_\alpha^C = u_\alpha^C = \frac{1}{4\pi\mu l}\left[K_1\left(\frac{r}{l}\right) - \frac{l}{r}\right]\frac{\varepsilon_{\alpha\gamma}x_\gamma}{r} \tag{205}$$

$$\Omega^C = \omega^C = \frac{1}{8\pi\mu l^2}K_0\left(\frac{r}{l}\right) \tag{206}$$

$$\Sigma_{\beta\alpha}^C = \sigma_{\beta\alpha}^C =$$
$$-\frac{1}{4\pi l^2}\left[K_0\left(\frac{r}{l}\right) + \frac{2l}{r}K_1\left(\frac{r}{l}\right) - \frac{2l^2}{r^2}\right]\frac{\varepsilon_{\alpha\gamma}x_\gamma x_\beta + \varepsilon_{\beta\gamma}x_\gamma x_\alpha}{r^2} + \frac{1}{4\pi l^2}K_0\left(\frac{r}{l}\right)\varepsilon_{\alpha\beta} \tag{207}$$

$$\mathsf{M}_\alpha^C = \mu_\alpha^C = \frac{1}{2\pi d}K_1\left(\frac{r}{l}\right)\frac{\varepsilon_{\alpha\beta}x_\beta}{r} \tag{208}$$

$$T_\alpha^C = t_\alpha^{(n)C} =$$
$$-\frac{1}{4\pi l^2}\left[K_0\left(\frac{r}{l}\right) + \frac{2l}{r}K_1\left(\frac{r}{l}\right) - \frac{2l^2}{r^2}\right]\frac{\varepsilon_{\alpha\gamma}x_\gamma x_\beta n_\beta + \varepsilon_{\beta\gamma}x_\gamma x_\alpha n_\beta}{r^2} + \frac{1}{4\pi l^2}K_0\left(\frac{r}{l}\right)\varepsilon_{\alpha\beta}n_\beta \tag{209}$$

$$M^C = m^{(n)C} = -\frac{1}{2\pi d}K_1\left(\frac{r}{l}\right)\frac{x_\beta n_\beta}{r} \tag{210}$$

where $U_\alpha^C$, $\Sigma_{\beta\alpha}^C$, $\mathsf{M}_\alpha^C$ and $T_\alpha^C$ represent, respectively, the corresponding displacement, force-stress, couple-stress and force-traction at $x$ caused by a unit in-plane concentrated couple at the origin. Meanwhile, $\Omega^C$ and $M^C$ represent the respective corresponding rotation and moment-traction at $x$ due to this unit in-plane concentrated couple at the origin.

## 4. Conclusions

We have derived the three- and two-dimensional fundamental solutions for isotropic size-dependent couple stress elasticity, based upon the fully determinate couple-stress theory. Recall that this new theory resolves all of the difficulties present in previous attempts to construct a viable size-dependent elasticity. Furthermore, since in this theory, body couples do not appear in the constitutive relations and everything depends on only a single size-dependent material constant, all expressions for the fundamental solutions are elegantly consistent and quite useful in practice. In particular, these solutions can be used directly as influence functions to analyze infinite domain problems or as kernels in integral equations for numerical analysis. For example, a boundary element method for plane problems of couple stress elasticity is developed in [19], based upon these fundamental solutions. Future work will include the formulation of boundary element methods for linear elastic fracture mechanics and for three-dimensional couple stress problems.